\documentclass[aps,prl,twocolumn,groupedaddress]{revtex4}

\usepackage{graphicx} 
\usepackage{dcolumn}
\usepackage{bm}

\begin{document}
\title{Fine structure in the tunneling spectra of electron-doped
cuprates: No coupling to magnetic resonance mode } 
\author{Guo-meng Zhao$^{1,2,*}$} 
\affiliation{$^{1}$Department of Physics and Astronomy, 
California State University, Los Angeles, CA 90032, USA~\\
$^{2}$Department of Physics, Faculty of Science, Ningbo
University, Ningbo, P. R. China}

\begin{abstract}
We analyze high-resolution
scanning tunneling spectra of  the electron-doped cuprate
Pr$_{0.88}$LaCe$_{0.12}$CuO$_{4}$ ($T_{c}$ = 24 K). We find that the spectral fine structure below 35 meV 
is consistent with strong coupling to a 
bosonic mode at 16 meV, in quantitative agreement with early tunneling
spectra of
Nd$_{1.85}$Ce$_{0.15}$CuO$_{4}$.  Since the energy of the bosonic mode
is significantly higher than that (9.5-11 meV) of the magnetic resonance-like
mode observed by inelastic neutron scattering, the coupling
feature at 16 meV cannot arise from strong coupling
to the magnetic mode.
The present work
thus demonstrates that 
the magnetic resonance-like mode cannot be the origin of high-temperature superconductivity 
in electron-doped cuprates.

\end{abstract}
\maketitle 

The microscopic pairing mechanism for high-temperature superconductivity
in cuprates remains elusive despite tremendous efforts for over twenty
years. The most central issue is the origin of the bosonic modes 
mediating the electron pairing. Most workers believe
that magnetic resonance modes, which have been observed in various
hole-doped double-layer cuprate 
systems, predominantly mediate the electron pairing. Recent
observation of a magnetic resonance-like mode at 9.5-11.0 meV in two
electron-doped cuprates \cite{Wilson,JZhao}
seems to suggest that the magnetic resonance is a universal property
of all cuprate systems and thus essential to the pairing mechanism of 
high-temperature superconductivity. However, this speculated magnetic pairing mechanism
is seriously undermined by recent optical experiments
\cite{YLee} which showed that the
electron-boson spectral function 
$\alpha{^2}(\omega)$$F(\omega)$ is independent of magnetic field, in
contradiction with the theoretical prediction based on the magnetic
pairing mechanism (see Fig.~9 of Ref.~\cite{YLee}). 
In contrast, 
extensive studies of various unconventional oxygen-isotope effects in 
hole-doped cuprates
have clearly shown strong electron-phonon interactions and the 
existence of polarons 
\cite{ZhaoAF,ZhaoYBCO,ZhaoLSCO,ZhaoNature97,ZhaoJPCM,Zhaoreview1,Zhaoreview2,Zhaoisotope,Keller1}. 
Neutron scattering 
\cite{McQueeney}, angle-resolved photoemission (ARPES) 
\cite{Lanzara01,Ros}, and Raman scattering \cite{Mis} experiments have
also demonstrated strong electron-phonon 
coupling. Further, ARPES data \cite{Zhou04} and  tunneling spectra
\cite{Ved,Shim,Gonnelli,Zhao07,Boz08} have consistently
provided direct evidence for strong coupling to multiple-phonon modes 
in hole-doped cuprates.
Therefore,  electron-phonon coupling in hole-doped cuprates should
play an important role in the pairing mechanism.

On the other hand, the role of electron-phonon coupling in the pairing
mechanism of electron-doped cuprates has not been clearly demonstrated.
Early tunneling spectra in Nd$_{1.85}$Ce$_{0.15}$CuO$_{4}$ (NCCO)
suggested predominantly phonon-mediated pairing \cite{Huang} while the
oxygen-isotope exponent $\alpha_{O}$ in Pr$_{1.85}$Ce$_{0.15}$CuO$_{4}$ 
was found to be 0.08$\pm$0.01 (Ref.~\cite{ZhaoPCCO}), which is significantly
below 0.5, expected for the phonon-mediated mechanism. Moreover, surface-sensitive ARPES 
experiment showed very weak electron-phonon coupling \cite{Arm03}, which may support
an alternative mechanism where the 10 meV magnetic resonance mode is mainly responsible for the pairing. 
If this were the case, the strong coupling to
the 10 meV magnetic excitation would show up in single-particle tunneling
microscopy that has a much higher energy resolution than ARPES. 
However, the early tunneling spectra \cite{Huang} do not seem to show 
this coupling feature at about 10 meV. One
might argue that the obsence of this coupling feature in the early
data could be due to a low experimental resolution and poor sample
quality.  
Therefore, it is essential to obtain
reproducible high-resolution  single-particle tunneling spectra and
analyze the spectra in a correct way to unambiguously address this issue.

Here we re-analyze high-resolution  single-particle tunneling spectra of the electron-doped cuprate
Pr$_{0.88}$LaCe$_{0.12}$CuO$_{4}$ ($T_{c}$ = 24 K) \cite{Nie}. The d$^{2}I$/d$V^{2}$ spectra 
reveal one dip and two peak features below $V$ = 35 mV, where $I$ is the tunneling current and $V$ is
the bias voltage.  We find that
these fine features are consistent with strong coupling to a 
bosonic mode at 16 meV, in quantitative agreement with early tunneling
spectra \cite{Huang} of
Nd$_{1.85}$Ce$_{0.15}$CuO$_{4}$.  Since the energy of the bosonic mode
is significantly higher than that (9.5-11 meV) of the magnetic resonance-like mode 
observed by inelastic neutron scattering \cite{Wilson,JZhao},
this coupling feature cannot arise from strong coupling
to the magnetic mode.
The present work
thus demonstrates that 
the magnetic resonance-like mode cannot be the origin of high-temperature superconductivity 
in electron-doped cuprates.

For conventional superconductors, the energies  of the 
phonon 
modes coupled to electrons can be precisely determined from the second derivative
tunneling 
spectra d$^{2}I$/d$V^{2}$. Measured from the isotropic $s$-wave 
superconducting gap $\Delta$, the energy positions of the dips
(minima) in d$^{2}I$/d$V^{2}$ correspond 
to those of the peaks in the electron-phonon spectral function 
$\alpha{^2}(\omega)$$F(\omega)$ (see Fig.~1 below and also
Refs.~\cite{MR,Carb}).  In a recent article 
\cite{Lee} attempting to show an important role of phonons in the 
electron pairing, the authors assign the energy (52 meV) of a peak position  in d$^{2}I$/d$V^{2}$ 
spectra of 
Bi$_{2}$Sr$_{2}$CaCu$_{2}$O$_{8+\delta}$ to the energy of a phonon 
mode.  Such an assignment is incorrect because the energies of phonon 
modes are equal to the energies of dip positions rather than peak 
positions in d$^{2}I$/d$V^{2}$ (see Fig.~1 below and also
Refs.~\cite{Zhao07,Boz08}). The same
mistake occurs in a more recent article \cite{Nie} where the authors also assign the energy
(10.5 meV) of a peak position  in d$^{2}I$/d$V^{2}$ spectra of 
Pr$_{0.88}$LaCe$_{0.12}$CuO$_{4}$ (PLCCO) to the energy of a bosonic mode. Since this mistakenly
assigned mode
energy (10.5 meV) is very close to the energy (9.5-11 meV) of the magnetic
resonance-like mode measured by inelastic neutron scattering
\cite{Wilson,JZhao}, the authors \cite{Nie} conclude that the 
magnetic resonance 
mode mediates electron pairing in electron-doped cuprates.

\begin{figure}[htb]
    \includegraphics[height=6.2cm]{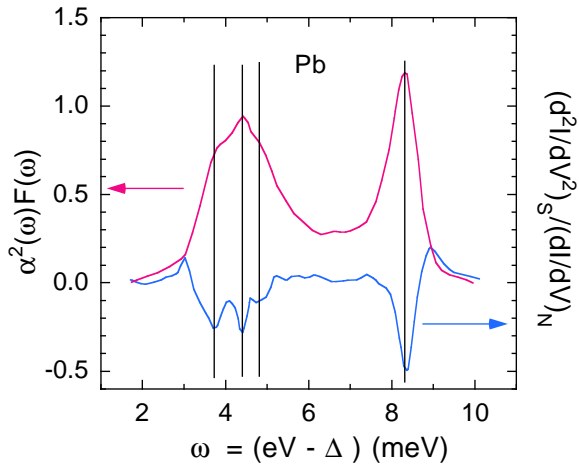}
 \caption[~]{Normalized second derivative
(d$^{2}I$/d$V^{2}$)$_{S}$/(d$I$/d$V$)$_{N}$ for the conventional
superconductor Pb along with the electron-phonon spectral function 
$\alpha{^2}(\omega)$$F(\omega)$ (where $S$ represents the
superconducting state and $N$ the normal state). The figure is reproduced from
Ref.~\cite{MR}. It is apparent that the dip features in (d$^{2}I$/d$V^{2}$)$_{S}$
match precisely with the peak features in $\alpha{^2}(\omega)$$F(\omega)$
(see the vertical solid lines). }
\end{figure}

Figure~1 shows the normalized second derivative
(d$^{2}I$/d$V^{2}$)$_{S}$/(d$I$/d$V$)$_{N}$ for the conventional
superconductor Pb along with the electron-phonon spectral function 
$\alpha{^2}(\omega)$$F(\omega)$ (where $S$ represents the
superconducting state and $N$ the normal state). The figure is reproduced from
Ref.~\cite{MR}. It is apparent that the dip features in (d$^{2}I$/d$V^{2}$)$_{S}$
match precisely with the peak features in $\alpha{^2}(\omega)$$F(\omega)$
(see the vertical solid lines). Therefore, the energy positions of bosonic
modes strongly coupled to electrons correspond to the energy
positions of the dip features (rather than the peak features) in (d$^{2}I$/d$V^{2}$)$_{S}$.

\begin{figure}[htb]
    \includegraphics[height=6cm]{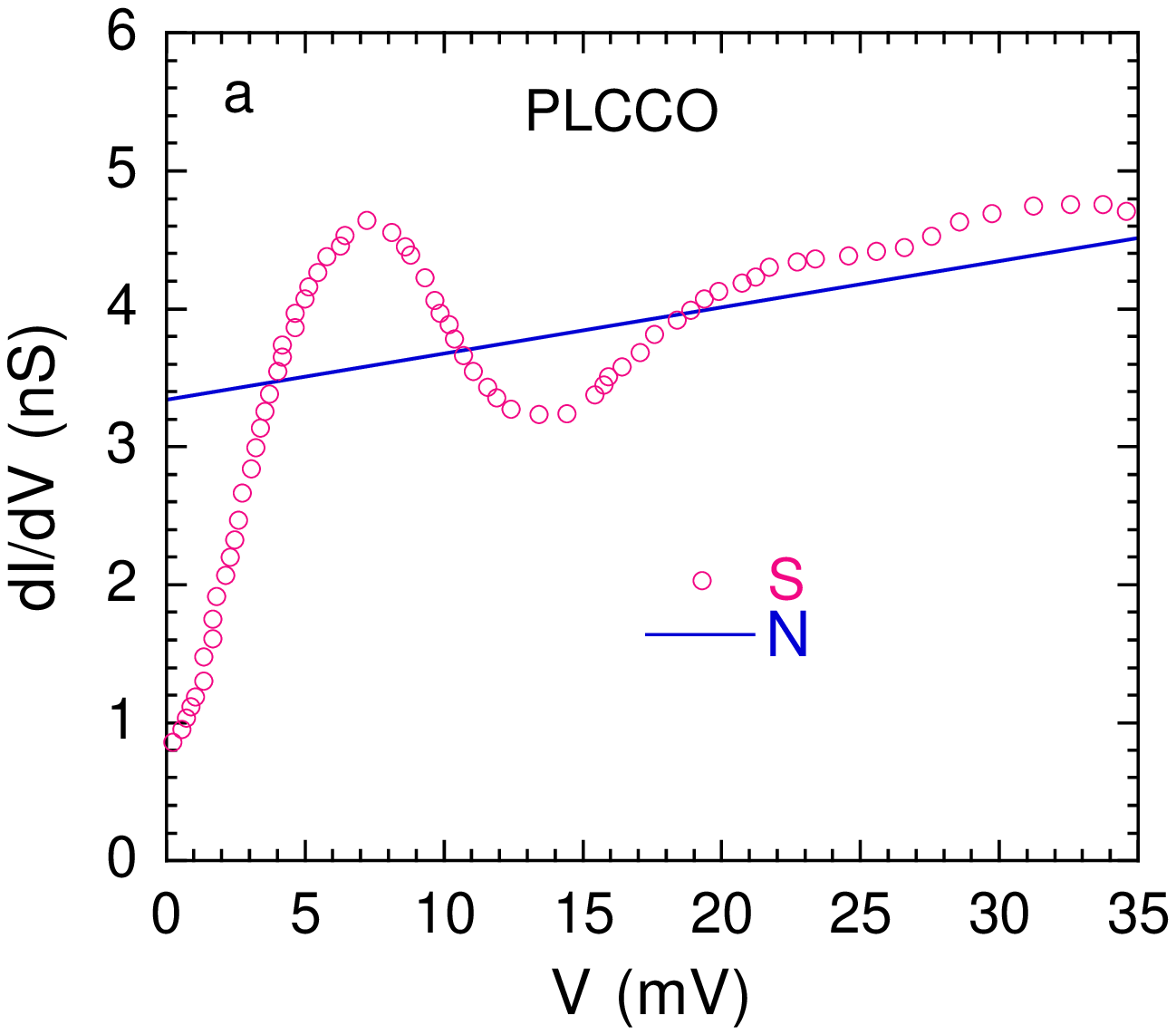}
    \includegraphics[height=6cm]{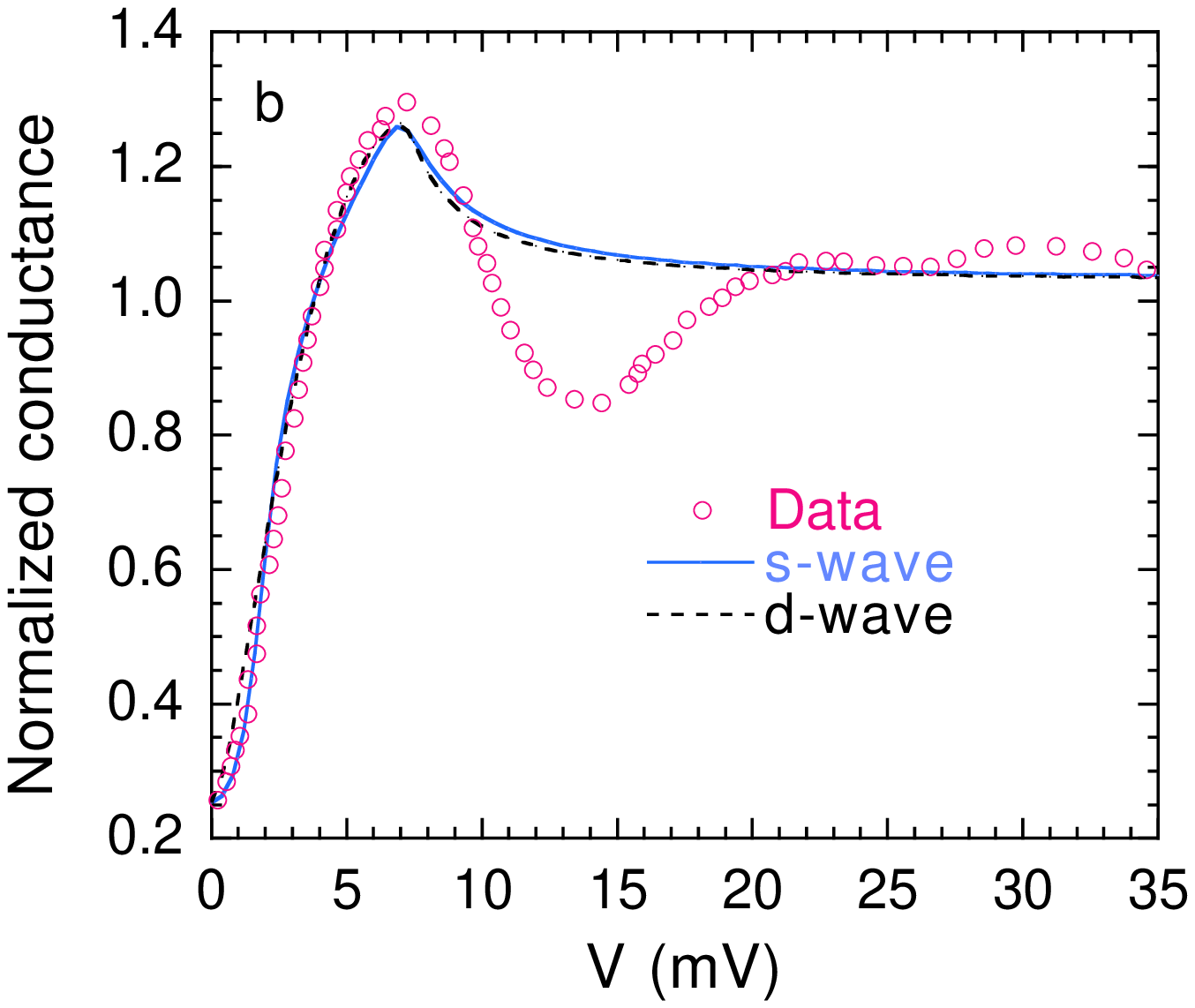}
	\caption[~]{a) Tunneling conductance (d$I$/d$V$)$_{S}$ in the
superconducting state for the
electron-doped Pr$_{0.88}$LaCe$_{0.12}$CuO$_{4}$ (PLCCO) crystal (open circles). The data are
digitized from Ref.~\cite{Nie}. The normal-state tunneling conductance
(d$I$/d$V$)$_{N}$ is
approximated by a straight line, which is obtained with  
conservation of states. b) Normalized conductance
(d$I$/d$V$)$_{S}$/(d$I$/d$V$)$_{N}$. The solid line is the
numerically calculated curve in terms of an anisotropic $s$-wave gap
symmetry and the dashed line  is the
numerically calculated curve in terms of  $d$-wave gap
symmetry.  }
\end{figure}

In Fig.~2a, we show the tunneling conductance (d$I$/d$V$)$_{S}$ in the
superconducting state for the
electron-doped Pr$_{0.88}$LaCe$_{0.12}$CuO$_{4}$ crystal. The data are
digitized from Ref.~\cite{Nie}. The normal-state tunneling conductance
(d$I$/d$V$)$_{N}$ is
approximated by a straight line, which is obtained by  
conservation of states, i.e., the superconducting spectral
deviation above the line is balanced
by the deviation below. Then, we obtain the normalized conductance
(d$I$/d$V$)$_{S}$/(d$I$/d$V$)$_{N}$, which is shown in Fig.~2b.

For an anisotropic gap function $\Delta (\theta)$, 
the directional dependence of the differential tunneling conduction is 
given by \cite{Suzuki}
\begin{equation}
\frac{dI}{dV} \propto \int_{0}^{2\pi}p(\theta 
-\theta_{0}) Re[\frac{eV - i\Gamma}{\sqrt{(eV - i\Gamma)^{2} - 
\Delta^{2}(\theta)}}]N(\theta) d\theta,
\end{equation}
where $N(\theta)$ represents the anisotropy of the band dispersion, 
$\Gamma$ is the life-time broadening parameter of an electron,
$p(\theta -\theta_{0})$ is the angle dependence of the 
tunneling probability and equal to $\exp [-\beta \sin^{2}(\theta -\theta_{0})]$, 
and $\theta_{0}$ is the angle of the tunneling 
barrier direction measured from the Cu-O bonding direction. For simplicity, we assume a cylindrical 
Fermi surface so that both $N(\theta)$ and $\beta$ are independent of 
the angle.  The solid line is the
numerically calculated curve using $\Gamma$ = 0.73 meV, $\beta$ = 6, $\theta_{0}$ =
0.18$\pi$, and  an $s$-wave gap function: $\Delta$ = $4.4 (1.0 - 0.6\sin
4\theta)$ meV. The dashed line  is the
numerically calculated curve using $\Gamma$ = 0.40 meV, $\beta$ = 7, $\theta_{0}$ =
$\pi$/4, and a simple $d$-wave gap function:  $\Delta$ = $7.2\cos
2\theta$ meV. It is interesting that the calculated
curves for the anisotropic
$s$-wave and $d$-wave gaps are all in good agreement with the data. We 
further find that 
isotropic $s$-wave gap is
inconsistent with the data. Therefore, the tunneling spectrum alone
rules out isotropic $s$-wave gap symmetry on the top surface of the
crystal but cannot make
distinction between $d$-wave and anisotropic $s$-wave gap symmetry.

\begin{figure}[htb]
     \vspace{-0.2cm}
    \includegraphics[height=13cm]{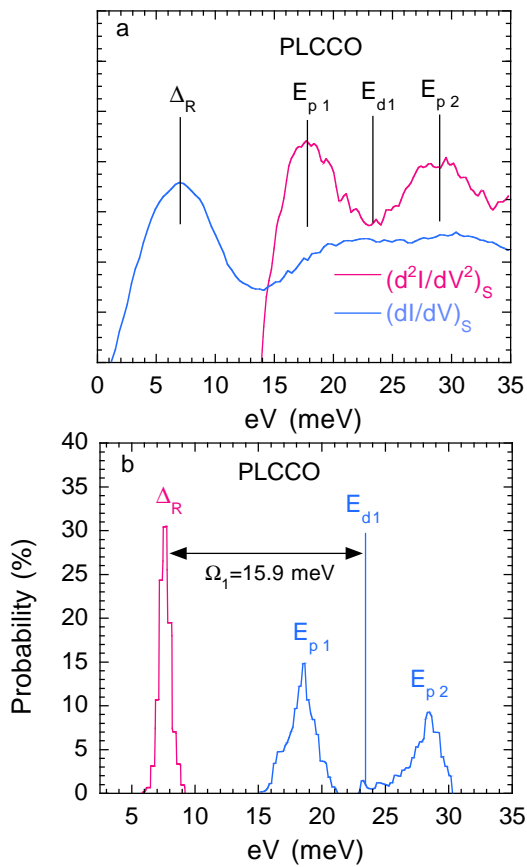}
    \vspace{-1cm}
	\caption[~]{a) The second derivative spectrum (d$^{2}I$/d$V^{2}$)$_{S}$
together with the
first derivative spectrum (d$I$/d$V$)$_{S}$ for the  Pr$_{0.88}$LaCe$_{0.12}$CuO$_{4}$ crystal.
The figure is reproduced from Ref.~\cite{Nie}.  b) A histogram of the occurrences of $\Delta_{R}$
(red) and the energies $E_{p1}$ and $E_{p2}$ (blue) for a map of
tunneling spectra on a 64~\AA$\times$64~\AA~area of the sample.  }
\end{figure}

In Fig.~3a, we show the second derivative spectrum (d$^{2}I$/d$V^{2}$)$_{S}$
together with the
first derivative spectrum (d$I$/d$V$)$_{S}$ for the  Pr$_{0.88}$LaCe$_{0.12}$CuO$_{4}$ crystal.
The figure is reproduced from Ref.~\cite{Nie}. In the (d$^{2}I$/d$V^{2}$)$_{S}$
spectrum,
there are two peak features at $E_{p1}$ = 17.8 meV and $E_{p2}$ = 29.0
meV, and one dip feature at $E_{d1}$ = 23.4 meV, as indicated in the
figure. It is clear that the dip feature is just half-way between the 
two peak features. The energy position $\Delta_{R}$ of the peak in the (d$I$/d$V$)$_{S}$
spectrum is about 7.0 meV. Following the result of Fig.~1 for Pb, the 
energy of the bosonic mode coupled strongly to electrons is
$\Omega_{1}$ = $E_{d1} -
\Delta_{R}$ = 16.4 meV, which is slightly lower than the energy of a very strong coupling
feature observed in hole-doped cuprates (e.g., 20 meV in Bi$_{2}$Sr$_{2}$CaCu$_{2}$O$_{8+\delta}$
\cite{Gonnelli,Zhao05,Zhao07} and  18 meV in La$_{2-x}$Sr$_{x}$CuO$_{4}$
\cite{Boz08}).

Figure~3b shows a histogram of the occurrences of $\Delta_{R}$
(red) and the energies $E_{p1}$ and $E_{p2}$ (blue) for a map of
tunneling spectra on a 64~\AA$\times$64~\AA~area of the sample.
According to the result in Fig.~3a, the
mid-point between $E_{p1}$ and $E_{p2}$ should mark $E_{d1}$.  Then, the difference between $E_{d1}$ and $\Delta_{R}$
is found to be 15.9 meV, that is,
$\Omega_{1}$ = 15.9 meV. More bosonic modes would be revealed if these
spectra were extended to higher energies.

Figure~4 shows electron-boson spectral functions  for two 
Nd$_{1.85}$Ce$_{0.15}$CuO$_{4}$ samples with $T_{c}$ = 22$\pm$2 K. The figure is reproduced from
Ref.~\cite{Huang}. The energies of the lowest bosonic
modes in the
spectral functions of the two crystals are 15.2 meV and 17.2 meV, respectively
(see arrows in the figure).
A simple average of the mode energies of the two NCCO samples is
16.2 meV, which 
is in quantitative agreement with the mode energy (15.9 meV) averaged 
from thousands of tunneling spectra of the PLCCO crystal \cite{Nie}. This quantitative
agreement clearly indicates that the strong coupling feature at about 
16 meV in the tunneling 
spectra of electron-doped cuprates is intrinsic.

\begin{figure}[htb]
    \includegraphics[height=6.2cm]{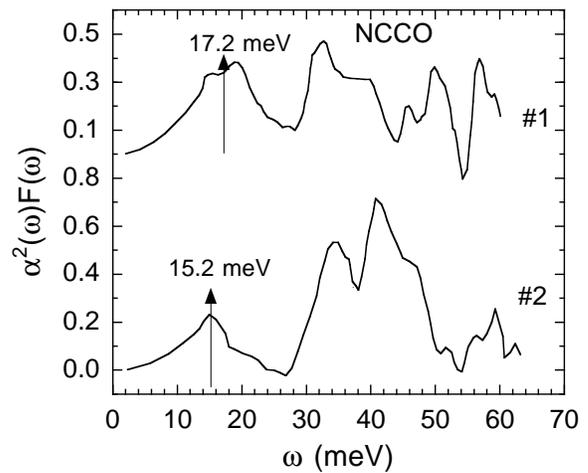}
	\caption[~]{Electron-boson spectral functions for two Nd$_{1.85}$Ce$_{0.15}$CuO$_{4}$ (NCCO)
samples with $T_{c}$ = 22$\pm$2 K.  The figure is reproduced from
Ref.~\cite{Huang}.}
\end{figure}
 
Having established the bosonic mode energy and statistics, we now discuss the nature of this 
mode. The measured mode energy of 16 meV rules out its
 connection to the magnetic resonance mode which has energy of 9.5 meV
 in NCCO (Ref.~\cite{JZhao}) and 11 meV in PLCCO (Ref.~\cite{Wilson}).
 Alternatively, this 16 meV bosonic mode should be associated with unscreened $c$-axis polar phonons
 (transverse optical phonons). Both ARPES data \cite{Mee1} of hole-doped
 Bi$_{2}$Sr$_{2}$CuO$_{6}$ and a theoretical study \cite{Mee2} indicate that the transverse optical (TO)
 phonons are strongly coupled to electrons due to the unscreened
 long-range interaction along the c-axis and play a predominant role
 in electron pairing. This long-range electron-phonon interaction 
 should be present in any layered system, but is often ignored when
 one theoretically calculates electron-phonon coupling.  For electron-doped 
 Pr$_{1.85}$Ce$_{0.15}$CuO$_{4}$, the lowest TO modes with 
 energies of 15.6
 meV ($E_{u}$ symmetry) and 17.0 meV ($A_{2u}$) were identified by
 infrared reflectivity measurements \cite{Crawford}. Since these two TO modes have
 energies very close to the energy of the bosonic mode seen in
 the tunneling spectra, it is natural that the 16 meV coupling 
 feature is
 associated with strong coupling of these TO phonon modes to
 electrons.

In summary, we have analyzed the high-resolution tunneling spectra of the electron-doped Pr$_{0.88}$LaCe$_{0.12}$CuO$_{4}$. We find 
that the spectral fine structure below 35 meV 
is consistent with strong coupling to a 
bosonic mode at 16 meV, in quantitative agreement with early tunneling
spectra \cite{Huang} of
Nd$_{1.85}$Ce$_{0.15}$CuO$_{4}$.  Since the energy of the bosonic mode
is significantly higher than that (9.5-11 meV) of the magnetic resonance-like
mode observed by inelastic neutron scattering, the coupling
feature at 16 meV cannot arise from strong coupling
to the magnetic mode.
The present work
thus demonstrates that 
the magnetic resonance-like mode cannot be the origin of high-temperature superconductivity 
in electron-doped cuprates.
~\\
$^{*}$ gzhao2@calstatela.edu

\bibliographystyle{prsty}

\end{document}